\documentclass[a4paper]{article}
\usepackage{listings}
\usepackage[svgnames]{xcolor}
\usepackage{multirow}
\usepackage{jheppub}  
\usepackage[sort&compress,numbers]{natbib}
\usepackage{dcolumn}
\usepackage[english]{babel}
\usepackage[utf8x]{inputenc}
\usepackage[T1]{fontenc}
\usepackage{palatino}
\usepackage{subfig}
\usepackage{amsmath}
\usepackage{afterpage}
\usepackage{graphicx, subcaption}
\usepackage[colorinlistoftodos]{todonotes}
\usepackage[colorlinks=true]{hyperref}
\usepackage{makeidx}
\usepackage{url}
\usepackage{mathptmx}
\usepackage[retainorgcmds]{IEEEtrantools}
\usepackage{mathtools, slashed}
\usepackage{physics}
\usepackage[compat=1.1.0]{tikz-feynhand}
\usepackage{tikz}
\newcommand{\be}{\begin{equation}}  
\newcommand{\ee}{\end{equation}}  
\newcommand{\bea}{\begin{eqnarray}}  
\newcommand{\eea}{\end{eqnarray}}  
\makeindex
\begin{document}

\vspace*{1.2cm}

\thispagestyle{empty}
\begin{center}
{\LARGE \bf Photon production from gluon splitting and fusion induced by a magnetic field in heavy-ion collisions}

\par\vspace*{7mm}\par

{

\bigskip

\large \bf Alejandro Ayala$^{1,2,3,*}$, Santiago Bernal-Langarica$^{1,\dagger}$, José Jorge Medina-Serna$^{1,\ddagger}$ and Ana Julia Mizher$^{2,4,\S}$}

\bigskip

{\large \bf  e-mail: *ayala@nucleares.unam.mx, ${}^\dagger$ santiago.bernal@correo.nucleares.unam.mx, ${}^\ddagger$ jose.medina@correo.nucleares.unam.mx, ${}^\S$ ana.mizher@unesp.br\\ *speaker}

\bigskip

{$^{1}$ Instituto de Ciencias Nucleares, Universidad Nacional Autónoma de México, Apartado Postal 70-543, CdMx 04510, Mexico\\
$^{2}$ Instituto de Física, Universidade de São Paulo, Rua do Matão, 1371, CEP 05508-090, São Paulo, SP, Brazil\\
$^{3}$ Instituto de Física Teórica, Universidade Estadual Paulista,
Rua Dr. Bento Teobaldo Ferraz, 271 - Bloco II, 01140-070 São Paulo, SP, Brazil\\
$^{4}$ Centro de Ciencias Exactas and Departamento de Ciencias Básicas,
Facultad de Ciencias, Universidad del Bío-Bío, Casilla 447, Chillán, Chile}

\bigskip

{\it Presented at the Workshop of Advances in QCD at the LHC and the EIC, CBPF, Rio de Janeiro, Brazil, November 9-15 2025}


\vspace*{15mm}

\end{center}
\vspace*{1mm}

\begin{abstract}
\phantom{} In heavy-ion collisions, an excess in photon production, together with a larger than expected positive elliptic flow, has been observed, a phenomenon commonly referred to as the direct photon puzzle. In this work we study the mechanism of photon production arising from gluon splitting and fusion during the pre-equilibrium stage in the presence of magnetic fields in peripheral heavy-ion collisions. We begin by analyzing the general tensor structure of the two-gluon one-photon vertex, computing it at the one-loop level for magnetic fields of arbitrary strength without resorting to additional approximations. Using these expressions, we calculate the contribution of gluon fusion and splitting to the photon yield, revealing that splitting dominates over fusion at low photon energies. Our results are compared with experimental data from the PHENIX collaboration. Finally, we incorporate a longitudinal anisotropy into the initial gluon distribution and find that it does not significantly alter the photon yield compared to an isotropic distribution.
\end{abstract}

 \section{Introduction}

 In the study of relativistic heavy-ion collisions, it has been proposed that very short-lived but high-intensity magnetic fields are produced in peripheral collisions at high energies~\cite{Skokov:2009qp}, with peak intensities as high as $B\sim 10^{19}$ G for RHIC energies~\cite{STAR:2019wlg,Brandenburg:2021lnj}. This has prompted efforts to find signals of its presence and its consequences in this environment, such as the possibility to link part of the excess yield and the strength of the elliptic flow coefficient $v_2$ of direct photons with effects associated with magnetic field induced processes. The large magnitude of the photons' $v_2$, which is comparable to that of hadrons~\cite{PHENIX:2011oxq, David:2019wpt}, together with its large yield, has been referred to as the \textit{direct photon puzzle}. For low photon's transverse momentum, the yield is dominated by thermal photons, produced at the very early thermal history of the collision. An early emission of direct photons appears to be confirmed by the dependence of $v_2$ on the photon's transverse momentum, which for large values is consistent with zero. Since photons are a colorless penetrating probe, that can only be boosted at their production times, then they should come from the early stages, where expansion velocities are small~\cite{David:2019wpt}.

 Several possible solutions to this photon puzzle have been proposed, including approaches based on hydrodynamical models~\cite{Gale:2021emg, Niemi:2015qia, Chatterjee:2011dw, Dasgupta:2019whr}, radiation coming from the preequilibrium stage~\cite{Monnai:2014kqa, Berges:2017eom, Oliva:2017pri, Garcia-Montero:2023lrd}, transport models~\cite{Xu:2004mz, Kasmaei:2019ofu}, intermediate semi-QGP states~\cite{Gale:2014dfa, Lee:2014pwa}, and others. These scenarios, reviewed in Refs.~\cite{David:2019wpt, Blau:2023bvi}, include magnetic fields as sources of electromagnetic radiation. These scenarios provide a natural anisotropic emission, contributing to the photon elliptic flow without requiring linking its strength to the flow properties of the system. 
 
 The presence of the magnetic field opens several possible channels for photon production. These include magnetic field-induced bremsstrahlung and pair annihilation in the QGP~\cite{Tuchin:2014iua, Wang:2020dsr, Wang:2022jxx}, electromagnetic radiation from the QED$\times$QCD conformal anomaly~\cite{Basar:2012bp}, photons radiated from $2\to 2$ scattering processes among quarks and gluons in weak magnetic fields~\cite{Sun:2024vsz}, holographic methods describing photon production from a strongly coupled plasma in the presence of a magnetic field~\cite{Muller:2013ila, Arciniega:2013dqa, Avila:2022cpa}, photon production due to stochastic fluctuations of magnetic fields~\cite{Castano-Yepes:2024vlj}, and gluon fusion and splitting during preequilibrium in the presence of strong magnetic fields~\cite{Ayala:2017vex, Ayala:2019jey, Ayala:2022zhu, Ayala:2024ucr}.
 
 While the amplitudes for gluon splitting and fusion are suppressed compared to quark or anti-quark splitting and quark--anti-quark annihilation amplitudes, which are the leading processes in the perturbative expansion, gluon fusion and splitting are enhanced compared to processes involving quarks. This is because, in preequilibrium, the occupation number of quarks is suppressed with respect to gluons by a factor of $\alpha_s ^2$~\cite{Baier:2000sb, Garcia-Montero:2019vju, Monnai:2019vup}. Moreover, by taking into account Pauli blocking in the final state, it is possible to see that processes involving quarks either in the initial or final state are overall suppressed, making gluon fusion/splitting an important channel for photon production. As previously stated,  Refs.~\cite{Ayala:2017vex, Ayala:2019jey} explored these gluon channels under the approximation of strong magnetic fields, which restricted their calculation to the low transverse momentum part of the spectrum. In addition, the approximation used in Ref.~\cite{Ayala:2022zhu} does not account for the full tensor structure with the symmetry properties to describe the matrix element for these processes. This shortcoming was corrected in Ref.~\cite{Ayala:2024ucr}, where the two-gluon one-photon vertex in the presence of a magnetic field was studied starting from first principles to find the correct tensor structure that describes this vertex. With that tensor structure in hand, the two-gluon one-photon vertex was computed at the one-loop order in the intermediate field regime, but the contribution of these processes to the photon yield was not computed. In this work, we present the contribution of the gluon fusion and splitting to the photon yield, for magnetic fields of arbitrary strength.

 This work is organized as follows: in Sec.~\ref{II} we review the general tensor structure of the two-gluon one-photon vertex in the presence of a constant magnetic field and particularize it for the case of on-shell gauge bosons. In Sec.~\ref{III} we present the computation of the one-loop approximation of this vertex for magnetic fields of arbitrary strength. These expressions are used in Sec.~\ref{IV} to show its contribution to the photons' yield and compare it with data from PHENIX. We finally summarize and conclude in Sec.~\ref{V}.


\section{General structure of the two-gluon one-photon vertex}\label{II}

The general structure of the two-gluon one-photon vertex in the presence of a magnetic field, which is represented in Fig.~\ref{fig:vertex}, was studied in Ref.~\cite{Ayala:2024ucr}. Here we summarize the most relevant aspects of its tensor structure. This vertex can be denoted as 
\begin{figure}
\centering
\begin{tikzpicture}
\begin{feynhand}
\setlength{\feynhandblobsize}{15mm}
\vertex (a) at (-2,2) {$a, \;\mu$}; \vertex  (b) at (-2,-2) {$b,\;\nu$};  \vertex[NEblob] (c) at (0,0) {B}; \vertex (d) at (2.5,0) {$\alpha$};
\propag [glu] (a) to [edge label' =$p_1$] (c);
\propag [glu] (b) to [edge label' =$p_2$] (c) ;
\propag [bos] (c) to [edge label =$q$] (d);
\end{feynhand}
\end{tikzpicture}
\caption{General representation of the two-gluon one-photon vertex. The shaded blob represents the effect of a magnetic field.}
\label{fig:vertex}
\end{figure}
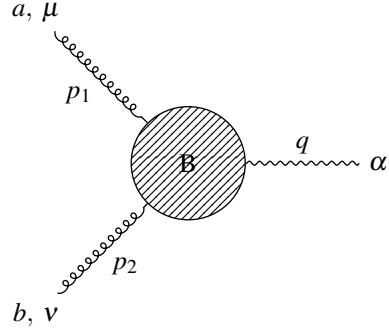
\begin{equation}
    \Gamma_{ab} ^{\mu\nu\alpha}(p_1, p_2, q),
\end{equation}
where $p_1$ and $p_2$ are the momentum of each gluon, $q$ is the momentum of the photon, $\mu$, $\nu$ and $\alpha$ are the Lorentz indices for the gluons and photon, respectively, and $a$ and $b$ the color indices of the gluons. Recall that, in the absence of a magnetic field, the vertex vanishes as required by Furry's theorem.

This vertex is a third rank tensor which is constrained by symmetry properties. First, gauge invariance requires the vertex to be transverse when contracted with the momentum of the external bosons. Second, since the gluons are indistinguishable, the vertex must be symmetric under the exchange of the gluon indices and momenta. Third, since charge ($C$) and parity ($P$) are good quantum numbers for the strong and electromagnetic interactions, and three neutral gauge bosons are involved, the vertex must be invariant under the combined action of the $CP$ transformation. All this properties must be translated into the properties of the chosen tensor basis and its coefficients to span the vertex. We introduce a set of four linearly independent vectors that span the four-dimensional Minkowski space-time and that also take into account for the presence of the external magnetic field. To start with, this vectors will be chosen as the photons momentum and the other three as general polarization vectors. We choose the Ritus base~\cite{Papanyan:1972cv, Papanyan:1974xa}, given by
\begin{IEEEeqnarray}{rCl}
\label{eq:basis}
q^\mu \nonumber \\
l_q ^\mu & \equiv & \hat{F} ^{\mu\beta} q_\beta \nonumber \\
l_q ^{*\mu} & \equiv & \hat{F} ^{*\mu\beta} q_\beta \nonumber \\
k_q ^\mu & \equiv & \frac{q^2}{l_q ^2} \hat{F}^{\mu\beta} \hat{F}_{\beta \sigma} q^\sigma + q^\mu,
\end{IEEEeqnarray}
where 
\begin{equation}
    \hat{F}^{\mu\beta} \equiv F^{\mu\beta} / |B|,
\end{equation}
with $F^{\mu\beta}$ the electromagnetic field strength tensor, $F^{*\mu\beta}$ its dual and $|B|$ the strength of the magnetic field. It can be seen that the three polarization vectors are transverse with respect to the photons' momentum. This set of vectors form  an orthogonal basis and satisfy the closure relation
\begin{equation}
    g^{\mu\nu} = \frac{q^\mu q^\nu}{q^2} + \frac{l_q ^\mu l_q ^\nu}{l_q ^2} + \frac{l_q ^{*\mu} l_q ^{*\nu}}{l_q ^{*2}} + \frac{k_q ^{*\mu} k_q ^{*\nu}}{k_q ^2}.
\end{equation}
Also, notice that for on-shell photons $q^2 = 0$, and thus the vector $k^\mu = q^\mu$, showing that the only two polarization vectors are $l_q ^\mu$ and $l_q ^{*\mu}$, as it is expected from the fact that for real photons there are only two polarizations. In the case of a constant magnetic field in the $\hat{z}$-direction, the explicit form of the normalized polarization vector of the Ritus basis is
\begin{IEEEeqnarray}{rCl}
    l_q ^\mu & = & \frac{1}{\sqrt{-q_\perp ^2}} (0, q_y, -q_x, 0) \nonumber \\
   l_q ^{*\mu} & = & \frac{1}{\sqrt{q_\parallel ^2}} (q_z, 0, 0, \omega_q) \nonumber \\ 
   l_q ^\mu & = & \frac{1}{\sqrt{q^2 q_\perp ^2 q_\parallel ^2}} (q_\perp ^2 \omega_q, - q_\parallel ^2 q_x, -q_\parallel^2 q_y, q_z),
\end{IEEEeqnarray}
where $q_\perp ^2 = -(q_x^2 + q_y ^2)$ and $q_\parallel ^2 = (\omega_q ^2 - q_z ^2)$.

At this point, we can repeat the previous procedure for the momentum of each gluon and thus obtain their corresponding polarization vectors, namely
\begin{IEEEeqnarray}{rCl}
    \text{photon } \alpha & \to & q^\alpha,\; l_q ^\alpha,\; l_q ^{*\alpha}, \; k_q ^{\alpha} \nonumber \\
    \text{gluon } \mu,\; a & \to & p_1 ^\mu,\; l_{p_1} ^\mu,\; l_{p_1} ^{*\mu}, \; k_{p_1} ^{\mu} \nonumber \\
    \text{gluon } \nu \; b & \to & p_2 ^\nu,\; l_{p_2} ^\nu,\; l_{p_2} ^{*\nu}, \; k_{p_2} ^{\nu}.
\end{IEEEeqnarray}
Therefore, the $i-$th basis element corresponds to one of the products of the three polarization vectors, where each factor is taken from one of the sets of polarization vectors for each particle. This corresponds to 27 different tensor structures, which can be reduced by imposing the symmetry restrictions previously stated. By imposing the symmetry under gluon exchange, the number of tensor structures reduces to 18. Now if we consider the properties of each structure under $C$ and $P$ and recall that the vertex must be invariant under $CP$, then there are four Lorentz scalars with definite properties under $C$ and $P$. Expressing any of the momentum vectors of the bosons as $p_m ^\mu$, with $m = 1$, 2, 3, then the available Lorentz scalars are
\begin{IEEEeqnarray}{rCl}
    S_{1mn} ^{++} & = & (p_m \cdot p_n) \nonumber \\
    S_{2mn} ^{++} & = & (p_m \cdot p_n)_\perp \nonumber \\
    S_{3mn} ^{-+} & = & p_m ^\mu \hat{F}_{\mu\nu} p_n ^\nu \nonumber \\
    S_{4mn} ^{--} & = & p_m ^\mu \hat{F}_{\mu\nu} ^* p_n ^\nu ,
\end{IEEEeqnarray}
where the notation $S_{jmn} ^{CP}$ refers to its properties under the $C$ and $P$ transformations.

Hitherto this discussion is valid for gauge bosons of arbitrary momenta, but from now on, we will restrict to the case of on-shell gauge bosons. As it was previously discussed, for on-shell bosons the $k_p$ polarization vector reduces to the boson momentum, and so all the terms involving it drop out and the number of terms is reduced to 5. An even further simplification is accomplished by noticing that for on-shell gauge bosons, conservation of energy-momentum
\begin{IEEEeqnarray}{rCl}
    \omega_q & = & \omega_{p_1} + \omega_{p_2} \nonumber \\
    \Vec{q} & = & \Vec{p}_1 + \Vec{p}_2,
\end{IEEEeqnarray}
requires that the gauge bosons are collinear and so, we can express the gluon momenta in terms of the photon's momentum. Another consequence is that when dealing with on-shell gauge bosons, only the scalar $S_{2mn} ^{++}$ is non-zero, and so, all Lorentz structures, odd with respect to either $C$ or $P$ are not available to express the on-shell vertex. Putting all this information together we arrive at
\begin{equation}
    \Gamma_{ab} ^{\mu\nu\alpha} (p_1, p_2, q)_{\text{on-shell}} = a_1 ^{++} \hat{l}_{p_1} ^\mu \hat{l}_{p_2} ^\nu \hat{l}_{q} ^\alpha + a_2 ^{++} \hat{l}_{p_1} ^{*\mu} \hat{l}_{p_2} ^{*\nu} \hat{l}_{q} ^\alpha + \frac{a_{10} ^{++}}{\sqrt{2}} \left( \hat{l}_{p_1} ^\mu \hat{l}_{p_2} ^{*\nu} + \hat{l}_{p_1} ^{*\mu} \hat{l}_{p_2} ^\nu \right) \hat{l}_{q} ^{*\alpha},
\end{equation}
which shows that only three independent tensor structures, together with their corresponding coefficients, are needed to span the on-shell two-gluon one-photon vertex. Given that the basis vectors are orthonormal, it is easy to find the coefficients $a_1 ^{++}$, $a_2 ^{++}$ and $a_{10} ^{++}$ by projecting the vertex $\Gamma_{ab} ^{\mu\nu\alpha}$ onto each of the tensors that make the basis.

\section{One-loop approximation for the two-gluon one-photon vertex for magnetic fields of arbitrary strength}\label{III}

\begin{figure}[t]
\centering
\begin{tikzpicture}
\begin{feynhand}
\vertex (a) at (0,0.75); \vertex  (b) at (0,-0.75);  \vertex (c) at (2,0.75); \vertex (d) at (2,-0.75); \vertex (e) at (3.5,0); \vertex (f) at (5,0); 
\vertex  (g) at (2,1) {$\mathcal{A}$};
\propag [glu] (a) to [edge label' =$p_1$] (c);
\propag [glu] (b) to [edge label' =$p_2$] (d) ;
\propag [fer] (d) to  (c);
\propag [fer] (c) to  (e);
\propag [fer] (e) to  (d);
\propag [bos] (e) to [edge label =$q$] (f);
\end{feynhand}
\end{tikzpicture}
\begin{tikzpicture}
\begin{feynhand}
\vertex (a) at (0,0.75); \vertex  (b) at (0,-0.75) {};  \vertex  (c) at (2,0.75) ;
\vertex  (g) at (2,1) {$\mathcal{B}$}; \vertex (d) at (2,-0.75); \vertex (e) at (3.5,0); \vertex (f) at (5,0); 
\propag [glu] (a) to [edge label' =$p_1$] (c);
\propag [glu] (b) to [edge label' =$p_2$] (d) ;
\propag [antfer] (d) to  (c);
\propag [antfer] (c) to  (e);
\propag [antfer] (e) to  (d);
\propag [bos] (e) to [edge label =$q$] (f);
\end{feynhand}
\end{tikzpicture}
\caption{One-loop diagrams contributing to the two-gluon one-photon vertex. Diagram $\mathcal{B}$ represents the charge conjugate of diagram $\mathcal{A}$. The four-momentum vectors are chosen such that $q=p_1+p_2$.}
\label{fig:Fusion}
\end{figure}
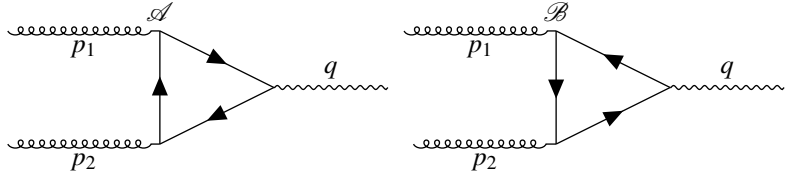

At leading order in the strong $\alpha_s$ and the electromagnetic $\alpha_{em}$ couplings, the vertex to be used to describe a scattering process involving two gluons and a photon, whether it is gluon fusion or splitting, is depicted in Fig.~\ref{fig:Fusion}. The amplitude corresponds to the sum of the Feynman diagrams represented as two fermion triangles with two gluons and one photon attached to the vertices and the charge in one and the other diagrams flowing in opposite directions. This amplitude was studied in detail in Ref.~\cite{Ayala:2024ucr} and the complete details can be found therein. Since the magnetic field breaks Lorentz symmetry, the vertex needs to be computed starting from configuration space. Each internal line corresponds to a fermion propagator in the presence of a magnetic field. The translationally invariant part of the propagator can be written, using Schwinger's proper time representation, as 
\begin{equation}
    S(p) = \int_0 ^\infty \frac{ds}{\cos(q_f Bs)} e^{is \left(p_\parallel ^2 + p_\perp ^2 \frac{\tan(q_f B s)}{q_f B s} - m_f ^2 + i \epsilon\right)} \left[ e^{iq_f B s \Sigma_3} (m_f + \slashed{p}_\parallel) + \frac{\slashed{p}_\perp}{\cos(q_f B s)}\right],
\end{equation}
where $m_f$ is the mass of the quark with flavor $f$ and $\Sigma_3 = i\gamma_1 \gamma_2$. The explicit form of the sum of the diagrams can be written as
\begin{IEEEeqnarray}{rCl}
    \Gamma_{ab} ^{\mu\nu\alpha} & = & -i g^2 q_f \int d^4x\, d^4 y\, d^4 z \int \frac{d^4 r_1}{(2\pi)^4}\frac{d^4 r_2}{(2\pi)^4}\frac{d^4 r_3}{(2\pi)^4} e^{ir_3\cdot(y-x)} e^{ir_2\cdot(x-z)}  e^{ir_1\cdot(z-y)}  e^{-ip_1\cdot z}  e^{-ip_2\cdot y}  e^{iq\cdot x} \nonumber \\
    & & \times \left\{ \Tr[\gamma_\alpha S(r_2) \gamma_\mu t_a S(r_1) \gamma_\nu t_b S(r_3)] \Phi(x,y,z,x) + \Tr[\gamma_\alpha S(r_3) \gamma_\nu t_b S(r_1) \gamma_\mu t_a S(r_2)] \Phi(x,z,y,x) \right\}, \IEEEeqnarraynumspace
\end{IEEEeqnarray}
where
\begin{IEEEeqnarray}{rCl}
    \Phi(x,y,z,x) & \equiv & \Phi(x,y) \Phi(y,z) \Phi(z,x) \nonumber \\
    \Phi(x,z,y,z) & \equiv & \Phi(x,z) \Phi(z,y) \Phi(y,x) = \Phi ^* (x,y,z,x),
\end{IEEEeqnarray}
are the product of Schwinger phases, defined as
\begin{equation}
    \Phi(x,x') \equiv \exp\left[iq_f \int_{x'} ^{x} d\xi^\mu \left[ A^\mu + \frac{1}{2}F_{\mu\nu} (\xi - x')^\nu \right]\right],
\end{equation}
with $q_f$ being the charge of the quark with flavor $f$, $g$ the quark gluon coupling, $t_c = \lambda_c / 2$, with $\lambda_c$ the Gell-Mann matrices. To describe a constant magnetic field in the $\hat{z}$-direction, the vector potential is written in the symmetric gauge, $A^\mu = \tfrac{B}{2}(0,-y,x,0)$.

After performing the integration over the space-time variables and the internal momentum components, we can introduce the following notation
\begin{IEEEeqnarray}{rCl}
    c_j & \equiv & \cos(q_f B s_j), \nonumber\\
    t_j & \equiv & \tan(q_f B s_j), \nonumber\\
    e_j & \equiv & c_j e^{i\text{sign}(q_fB) q_f B s_j \Sigma_3},
\end{IEEEeqnarray}
with $s_j$ being the Schwinger proper-time parameters associated with the internal fermion lines. After a lengthy but straightforward calculation, where we take the on-shell limit, we get
\begin{IEEEeqnarray}{rCl}
    \label{matelem}
    \Gamma^{\mu\nu\alpha}_{ab} & = & -i\frac{g^2q_f^2B\ }{(2\pi^2)}\Tr[t_at_b] \delta^4(p_1+p_2-q) \int_0^\infty\frac{ds_1ds_2ds_3}{c_1^2c_2^2c_3^2}\left(\frac{1}{t_1t_2t_3-t_1-t_2-t_3}\right)\left(\frac{e^{-ism_f^2}}{s}\right)\nonumber\\
    & & \times e^{-\frac{i}{s}\left(s_1s_3\omega_{p_1}^2+s_2s_3\omega_{p_2}^2+s_1s_2\omega_q^2\right)\frac{q_\perp^2}{\omega_q^2}} e^{-\frac{i}{\omega_q^2}\frac{q_\perp^2}{|q_fB|}\left(\frac{1}{t_1t_2t_3-t_1-t_2-t_3}\right)\left(t_1t_3\omega_{p_1}^2+t_2t_3\omega_{p_2}^2+t_1t_2\omega_q^2\right)}\nonumber\\
    & & \sum_{j=1}^{19}\left(T_{{\mathcal{A}}j}^{\mu\nu\alpha}+T_{{\mathcal{B}}j}^{\mu\nu\alpha}\right),
\end{IEEEeqnarray}
where $s = s_1 + s_2 + s_3$ and $T_{\mathcal{I}} ^{\mu\nu\alpha}$ are the traces over Dirac space corresponding to diagrams $\mathcal{A}$ and $\mathcal{B}$ of Fig.~\ref{fig:Fusion}, and that depend on the Schwinger parameters, as well as on the gauge bosons momenta. The explicit form of these traces can be found in Ref.~\cite{Ayala:2024ucr}.

\section{Contributions of gluon fusion and splitting to the photon yield}\label{IV}

The invariant momentum distribution of photons is given by
\begin{equation}
    \omega_q \dv{N^{\text{mag}}}{{}^3 q} = \frac{\mathcal{VT}}{2(2\pi)^3} \int \frac{d^3p_1}{(2\pi)^3 2p_1}\frac{d^3p_2}{(2\pi)^3 2p_2} n(p_1) n(p_2) \frac{1}{4} \sum_f |\mathcal{M}_f|^2,
\end{equation}
where $\mathcal{VT}$ represents the volume of the space-time region where photons are being produced, $n$ is the gluon distribution and $\mathcal{M}_f$ is the matrix element for the production of photons with the fermion $f$ as an intermediate state, which accounts for gluon fusion ($gg\to \gamma$ and splitting ($g\to g\gamma$), i.e., $|\mathcal{M}_f|^2 = |\mathcal{M}_{f, gg\to\gamma}|^2 + |\mathcal{M}_{f,g\to g\gamma}|^2$. Recall that this processes are related to each other by means of the crossing symmetry as $\mathcal{M}_{f,gg\to\gamma} = \mathcal{M}_{f,g\to g\gamma}$. Hence the matrix element can be written as
\begin{equation}
    \sum_f |\mathcal{M}_f|^2 = \sum_f \left[ (2\pi)^4 \delta^{(4)}(q - p_1 - p_2) |\Gamma_f|^2 + (2\pi)^4 \delta^{(4)}(q - p_1 + p_2) |\Gamma_f|^2 \right],
\end{equation}
with $\Gamma_f$ the matrix element of $gg\to\gamma$, studied in the previous section. Then, the photon yield for central rapidity is given by
\begin{equation}
    \frac{1}{\omega_q} \dv{N^{\text{mag}}}{\omega_q} = \frac{\mathcal{VT}}{8(2\pi)^5\omega_q} \int_0 ^{2\pi} d\phi \int_0 ^{\pi} d\theta \int dp_1\, n(p_1) \frac{1}{4} \sum_f \left[ n(q-p_1) |\Gamma_f|_{p_2 = q-p_1} ^2 + (1 + n(p-q_1)) |\Gamma_f| _{p_2 = p_1-q} ^2\right],
\end{equation}
where $\theta$ is the angle with respect to the magnetic field and $\phi$ is the azimuthal angle with respect to the beam. At this point it should be noticed that in a first approximation, the gluon distribution can be taken as an isotropic Bose-Einstein distribution from the shattered glasma~\cite{Ayala:2017vex}
\begin{equation}
    n(\omega) = \frac{\eta}{e^{\omega / \Lambda_s} - 1},
    \label{eq:cgc-dist}
\end{equation}
where $\eta$ and $\Lambda_s$ represent a high gluon occupation factor and saturation momentum scale, respectively. With this at hand, we can compute the contribution of these processes to the photon yield and compare it with the data from the PHENIX experiment.

\begin{figure}
    \centering
    \subfloat[$B = 3m_\pi^2$]{{\includegraphics[width=0.65\textwidth]{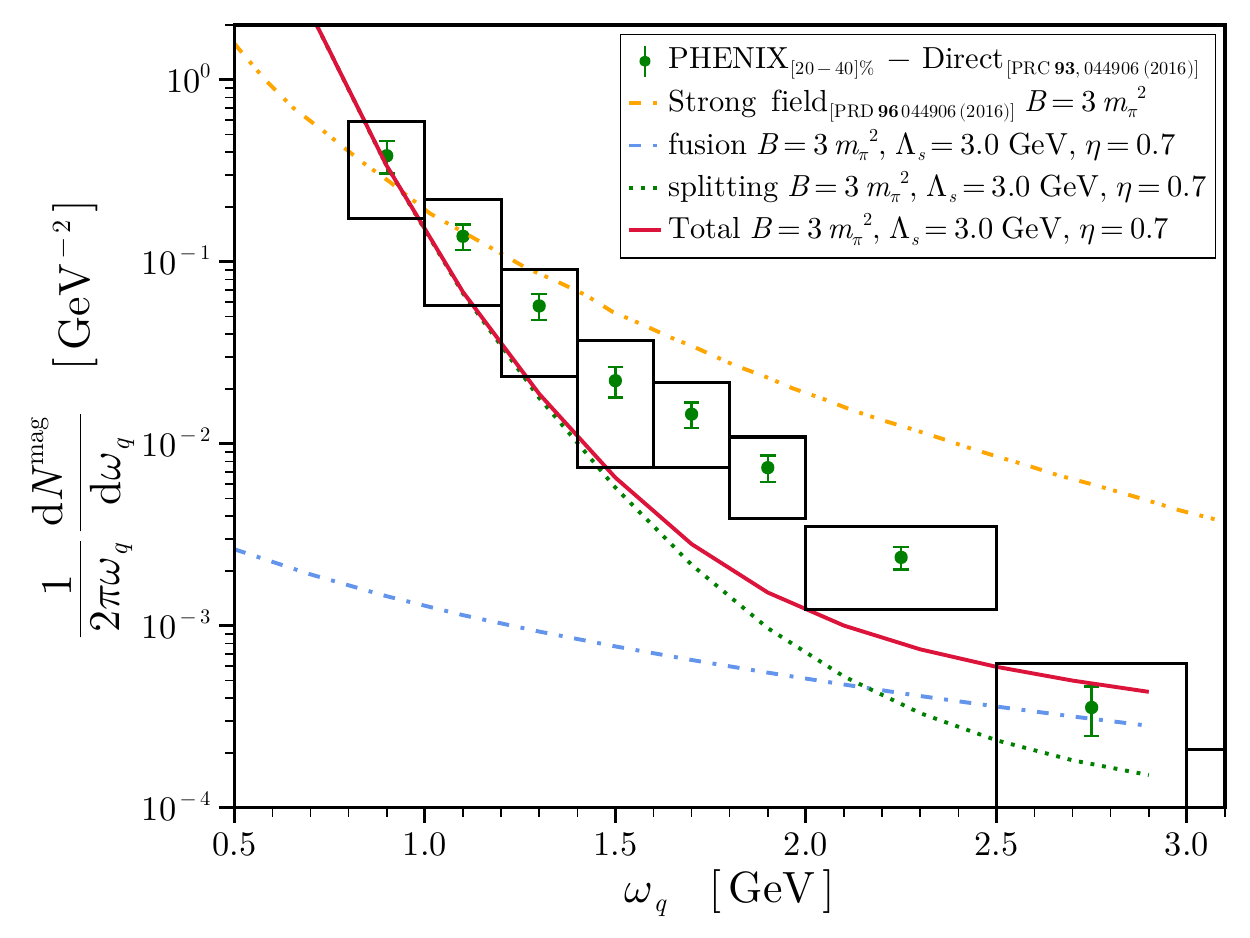}}} \quad
    \subfloat[$B = 10m_\pi^2$]{{\includegraphics[width=0.65\textwidth]{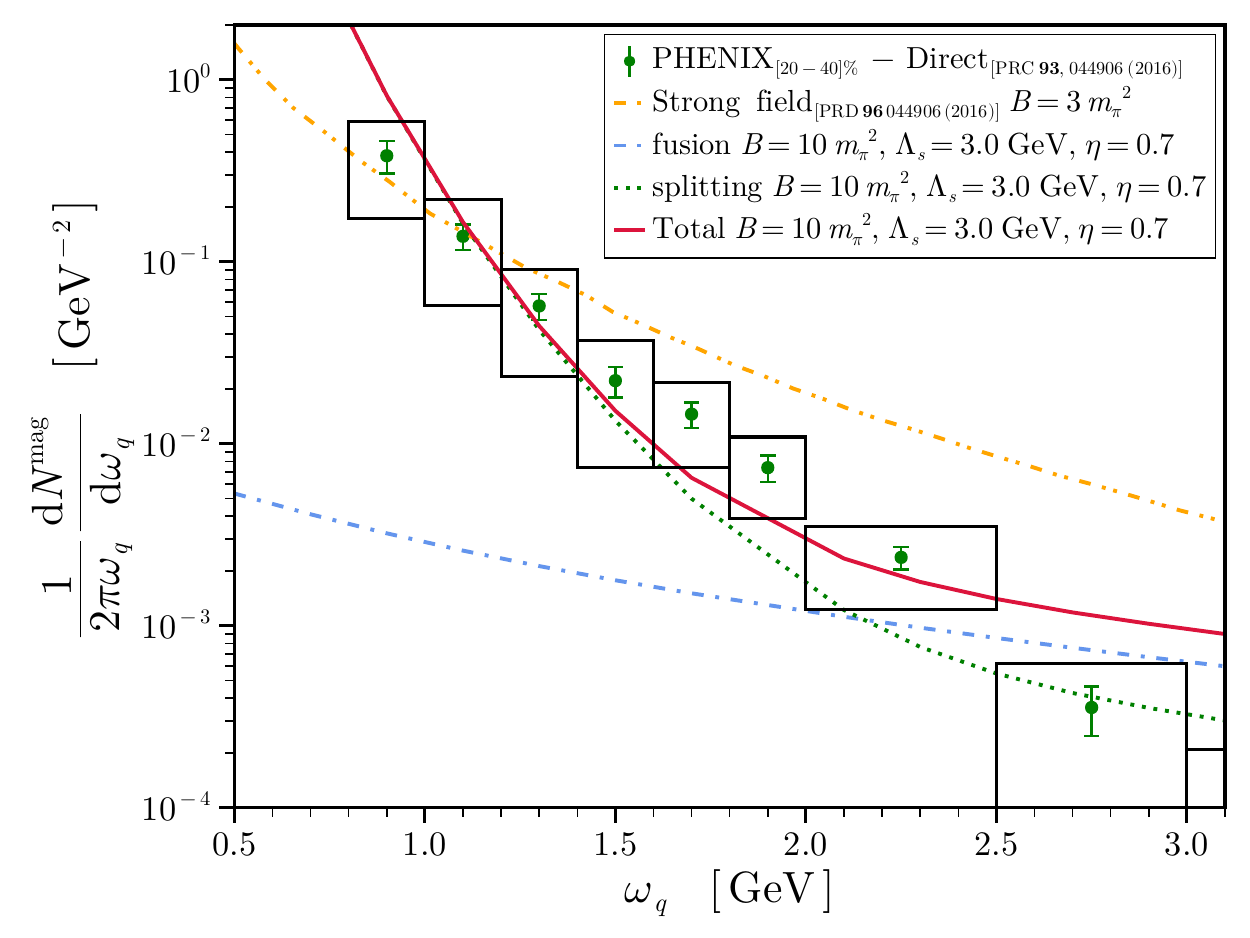}}}
    \caption{The contribution of gluon fusion (dashed-dotted blue line), splitting (dotted green line) and overall (solid red line), to the photon yield compared with the difference between PHENIX data~\cite{PHENIX:2022rsx} and the calculations of Ref.~\cite{Gale:2021emg} for the [20 - 30]\% centrality class (green markers), together with the strong field approximation of Ref.~\cite{Ayala:2017vex} (dashed-dotted yellow line). (a) shows the contribution for $B = 3m_\pi^2$ and (b) for $B = 10m_\pi^2$ }
    \label{Fig:iso_diff_B}
\end{figure}

\begin{figure}
    \centering
    \includegraphics[width=0.65\linewidth]{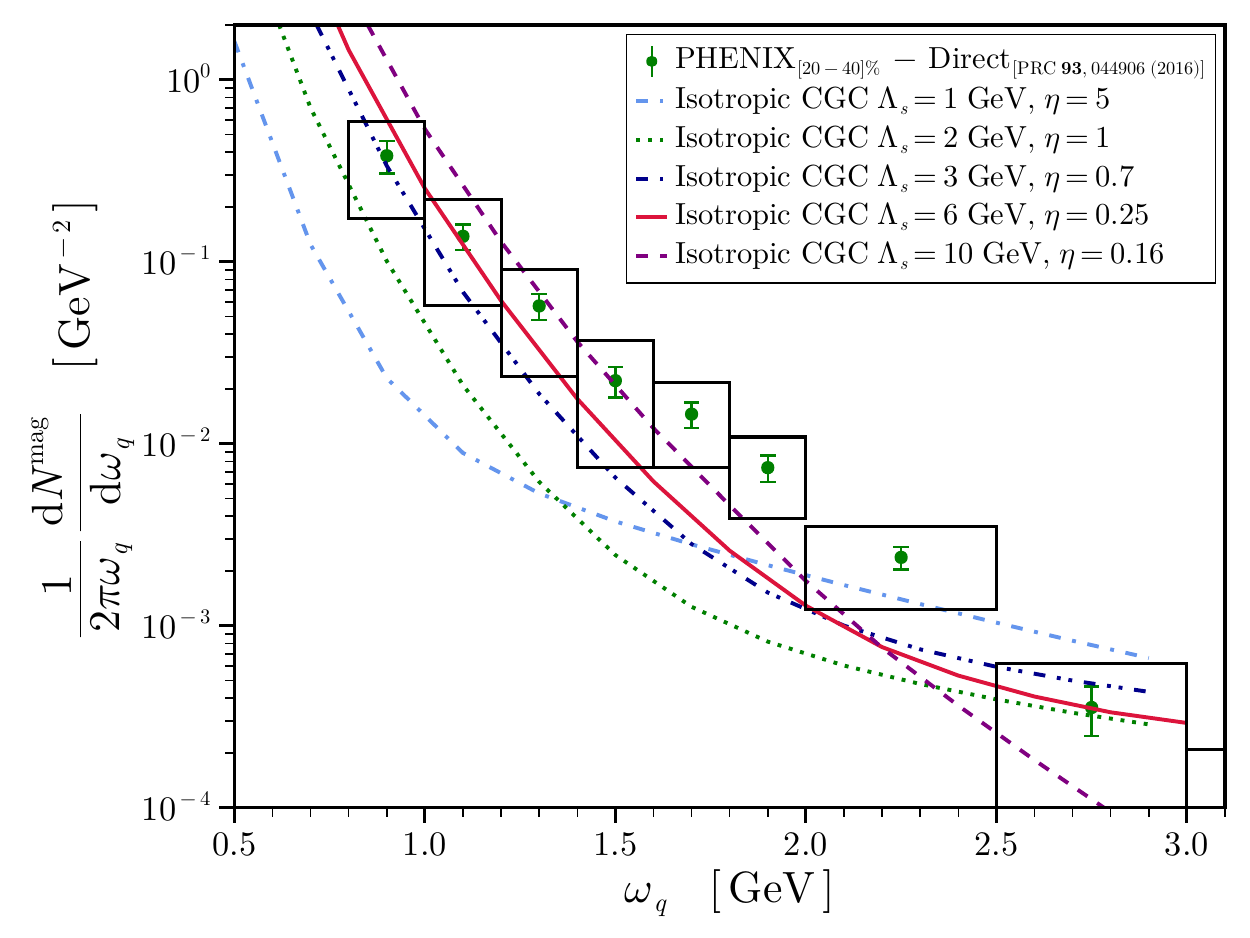}
    \caption{Effect of the gluon saturation scale, $\Lambda_s$, on the photon yield, compared with the difference between PHENIX data~\cite{PHENIX:2022rsx} and the calculations of Ref.~\cite{Gale:2021emg} for the [20 - 30]\% centrality class and for $B = 3m_\pi ^2$.}
    \label{Fig:iso_diff_lambdas}
\end{figure}

Figure~\ref{Fig:iso_diff_B} shows the difference between the PHENIX data~\cite{PHENIX:2022rsx} for the invariant momentum distribution and the state-of-the-art hydrodynamical calculations of Ref.~\cite{Gale:2021emg}, as well as the contribution of the gluon fusion, splitting, and the total contribution to the yield, for a couple of values of the magnetic field and the strong field approximation from Ref.~\cite{Ayala:2017vex}. It can be seen that the gluon splitting dominates over the fusion for small values of the photon energy. On the other hand, Fig.~\ref{Fig:iso_diff_lambdas} shows the effect of the gluon saturation scale on the yield. As it can be seen, it is precisely this parameter that controls the curvature of the yield.

\begin{figure}
    \centering
    \includegraphics[width=0.65\linewidth]{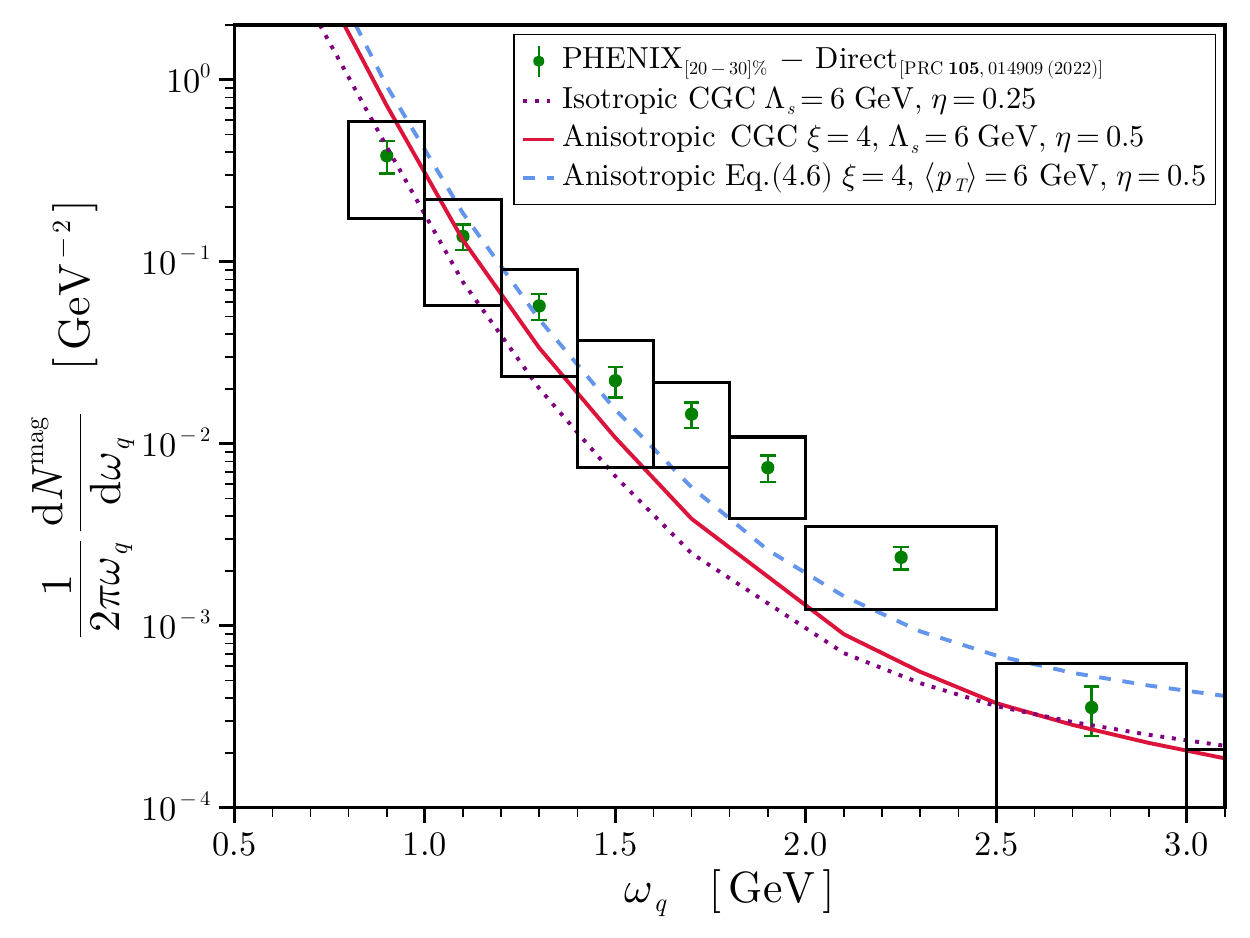}
    \caption{Comparison between the isotropic distribution of Eq.~\eqref{eq:cgc-dist} (dotted purple line) and the anisotropic distributions of Eq.~\eqref{eq:anis-cgc} (red solid line) and Eq.~\eqref{eq:dist_kurkela} (blue dashed line) and the PHENIX data~\cite{PHENIX:2022rsx} minus the calculations of Ref.~\cite{Gale:2021emg} (green markers) for $B = 10 m_\pi ^2$.}
    \label{fig:diff_dists}
\end{figure}

Recall that during the pre-equilibrium stage, the strong color fields dominate and undergo a significant anisotropic expansion in the longitudinal direction, which is also known as the beam direction. Consequently, the initial gluon occupation number can be expressed in terms of a distribution in momentum space that incorporates an anisotropy coefficient for the longitudinal momentum component~\cite{Kurkela:2015qoa}. Moreover, the magnetic field also generates an anisotropic pressure between the parallel and perpendicular components, relative to the magnetic field itself, which further contributes to the overall anisotropy observed during the pre-equilibrium phase~\cite{Ayala:2024jvc}. Therefore, this anisotropy must be included in the gluon distribution. In this work we included the anisotropy in the distribution of Eq.~\eqref{eq:cgc-dist} as
\begin{equation}
    n(p_L,p_\perp) = \frac{\eta}{e^{\sqrt{\xi^2 p_L ^2 + p_\perp ^2} / \Lambda_s} - 1},
    \label{eq:anis-cgc}
\end{equation}
where $\xi$ is the anisotropy coefficient for the longitudinal momentum component, $p_L$. Additionally, we also explored the anisotropic distribution for a color-glass condensate-inspired gluon-dominated initial state, which was used in Refs.~\cite{Kurkela:2015qoa, Garcia-Montero:2023lrd},
\begin{equation}
    n(p_L, p_\perp) ) = A \eta\langle p_\perp \rangle \frac{e^{-2(\xi^2 p_L^2 + p_\perp ^2) / 3 \langle p_\perp \rangle^2}}{\sqrt{\xi^2 p_L ^2 + p_\perp ^2}},
    \label{eq:dist_kurkela}
\end{equation}
where $\xi$ is the same as in Eq.~\eqref{eq:anis-cgc}, $A = 5.34$ is such that the comoving energy density is fixed~\cite{Garcia-Montero:2023lrd, Kurkela:2015qoa}, $\eta$ is the same as in Eq.~\eqref{eq:cgc-dist} and $\langle p_T \rangle$ is the mean transverse momentum. The comparison between the yields given by the anisotropic distributions of Eqs.~\eqref{eq:anis-cgc} and~\eqref{eq:dist_kurkela}, and the isotropic distribution of Eq.~\eqref{eq:cgc-dist}, is shown in Fig.~\ref{fig:diff_dists}. As it can be seen, the yield is not significantly affected by the momentum anisotropy and all of the studied distributions agree with the difference between the PHENIX and hydrodynamical calculations.

\section{Summary and conclusions}~\label{V}

In this work, we have presented the fundamental structure of the two-gluon one-photon vertex, derived from first principles and its symmetry properties, for gauge bosons of arbitrary momenta. We have further specialized these results for on-shell bosons and discovered that the vertex comprises only three distinct tensor structures. Notably, we computed this vertex up to the one-loop level without resorting to any additional approximations, enabling us to comprehensively analyze its contributions across a wide range of validity. The vertex was employed to investigate the contributions of the magnetic field-related channels of gluon fusion and splitting, with the latter dominating over the former at low energies. Additionally, we incorporated a longitudinal anisotropy into the gluon distribution, but found no significant deviations with respect to the isotropic distribution. We anticipate that this anisotropy will manifest in the computation of the elliptic flow, which is an ongoing research area that will be reported elsewhere.

\section*{Acknowledgments}

The authors thank J.-F. Paquet for kindly sharing their data. S. B.-L. and J. J. M.-S. each acknowledge the financial support of fellowships granted by Secretaría de Ciencia, Humanidades, Tecnología e Innovación as part of the Sistema Nacional de Posgrados. A. A. thanks the colleagues and staff of Universidade de São Paulo, of Instituto de Física Teórica, UNESP and of Universidade Cidade de São Paulo for their kind hospitality during a sabbatical stay. A. A. also acknowledges support from the PASPA program of the Dirección General de Asuntos del Personal Académico (DGAPA) of the Universidad Nacional Autónoma de México (UNAM) for the sabbatical stay during which this research was carried out. Support for this work was received in part via Secretaría de Ciencia, Humanidades, Tecnología e Innovación (SECIHTI), México Grants No. CF-2023-G-433 and No. CIORGANISMOS-2025-17. This study was financed, in part, by the São Paulo Research Foundation (FAPESP), Brazil. Process Numbers 2023/08826-7 and 2024/18493-8.

\bibliographystyle{apsrev4-1}
\bibliography{biblio}

@article{Skokov:2009qp,
    author = "Skokov, V. and Illarionov, A. Yu. and Toneev, V.",
    title = "{Estimate of the magnetic field strength in heavy-ion collisions}",
    eprint = "0907.1396",
    archivePrefix = "arXiv",
    primaryClass = "nucl-th",
    doi = "10.1142/S0217751X09047570",
    journal = "Int. J. Mod. Phys. A",
    volume = "24",
    pages = "5925--5932",
    year = "2009"
}

@article{STAR:2019wlg,
    author = "Adam, Jaroslav and others",
    collaboration = "STAR",
    title = "{Measurement of $e^+e^-$ Momentum and Angular Distributions from Linearly Polarized Photon Collisions}",
    eprint = "1910.12400",
    archivePrefix = "arXiv",
    primaryClass = "nucl-ex",
    doi = "10.1103/PhysRevLett.127.052302",
    journal = "Phys. Rev. Lett.",
    volume = "127",
    number = "5",
    pages = "052302",
    year = "2021"
}

@article{Brandenburg:2021lnj,
    author = "Brandenburg, James Daniel and Zha, Wangmei and Xu, Zhangbu",
    title = "{Mapping the electromagnetic fields of heavy-ion collisions with the Breit-Wheeler process}",
    eprint = "2103.16623",
    archivePrefix = "arXiv",
    primaryClass = "hep-ph",
    doi = "10.1140/epja/s10050-021-00595-5",
    journal = "Eur. Phys. J. A",
    volume = "57",
    number = "10",
    pages = "299",
    year = "2021"
}

@article{David:2019wpt,
    author = "David, Gabor",
    title = "{Direct real photons in relativistic heavy ion collisions}",
    eprint = "1907.08893",
    archivePrefix = "arXiv",
    primaryClass = "nucl-ex",
    doi = "10.1088/1361-6633/ab6f57",
    journal = "Rept. Prog. Phys.",
    volume = "83",
    number = "4",
    pages = "046301",
    year = "2020"
}

@article{PHENIX:2011oxq,
    author = "Adare, A. and others",
    collaboration = "PHENIX",
    title = "{Observation of direct-photon collective flow in $\sqrt{s_{NN}}=200$ GeV Au+Au collisions}",
    eprint = "1105.4126",
    archivePrefix = "arXiv",
    primaryClass = "nucl-ex",
    doi = "10.1103/PhysRevLett.109.122302",
    journal = "Phys. Rev. Lett.",
    volume = "109",
    pages = "122302",
    year = "2012"
}

@article{Gale:2021emg,
    author = {Gale, Charles and Paquet, Jean-Fran{\c{c}}ois and Schenke, Bj{\"o}rn and Shen, Chun},
    title = "{Multimessenger heavy-ion collision physics}",
    eprint = "2106.11216",
    archivePrefix = "arXiv",
    primaryClass = "nucl-th",
    doi = "10.1103/PhysRevC.105.014909",
    journal = "Phys. Rev. C",
    volume = "105",
    number = "1",
    pages = "014909",
    year = "2022"
}

@article{Niemi:2015qia,
    author = "Niemi, H. and Eskola, K. J. and Paatelainen, R.",
    title = "{Event-by-event fluctuations in a perturbative QCD + saturation + hydrodynamics model: Determining QCD matter shear viscosity in ultrarelativistic heavy-ion collisions}",
    eprint = "1505.02677",
    archivePrefix = "arXiv",
    primaryClass = "hep-ph",
    doi = "10.1103/PhysRevC.93.024907",
    journal = "Phys. Rev. C",
    volume = "93",
    number = "2",
    pages = "024907",
    year = "2016"
}

@article{Chatterjee:2011dw,
    author = "Chatterjee, Rupa and Holopainen, Hannu and Renk, Thorsten and Eskola, Kari J.",
    title = "{Enhancement of thermal photon production in event-by-event hydrodynamics}",
    eprint = "1102.4706",
    archivePrefix = "arXiv",
    primaryClass = "hep-ph",
    doi = "10.1103/PhysRevC.83.054908",
    journal = "Phys. Rev. C",
    volume = "83",
    pages = "054908",
    year = "2011"
}

@article{Dasgupta:2019whr,
    author = "Dasgupta, Pingal and Chatterjee, Rupa and Srivastava, Dinesh K.",
    title = "{Directed flow of photons in Cu+Au collisions at RHIC}",
    eprint = "1901.04943",
    archivePrefix = "arXiv",
    primaryClass = "nucl-th",
    doi = "10.1088/1361-6471/ab920e",
    journal = "J. Phys. G",
    volume = "47",
    number = "8",
    pages = "085101",
    year = "2020"
}

@article{Blau:2023bvi,
    author = "Blau, Dmitry and Peresunko, Dmitri",
    title = "{Direct Photon Production in Heavy-Ion Collisions: Theory and Experiment}",
    doi = "10.3390/particles6010009",
    journal = "Particles",
    volume = "6",
    number = "1",
    pages = "173--187",
    year = "2023"
}

@article{Monnai:2014kqa,
    author = "Monnai, Akihiko",
    title = "{Thermal photon $v_2$ with slow quark chemical equilibration}",
    eprint = "1403.4225",
    archivePrefix = "arXiv",
    primaryClass = "nucl-th",
    doi = "10.1103/PhysRevC.90.021901",
    journal = "Phys. Rev. C",
    volume = "90",
    number = "2",
    pages = "021901",
    year = "2014"
}

@article{Berges:2017eom,
    author = "Berges, Jurgen and Reygers, Klaus and Tanji, Naoto and Venugopalan, Raju",
    title = "{Parametric estimate of the relative photon yields from the glasma and the quark-gluon plasma in heavy-ion collisions}",
    eprint = "1701.05064",
    archivePrefix = "arXiv",
    primaryClass = "nucl-th",
    doi = "10.1103/PhysRevC.95.054904",
    journal = "Phys. Rev. C",
    volume = "95",
    number = "5",
    pages = "054904",
    year = "2017"
}

@article{Oliva:2017pri,
    author = "Oliva, L. and Ruggieri, M. and Plumari, S. and Scardina, F. and Peng, G. X. and Greco, V.",
    title = "{Photons from the Early Stages of Relativistic Heavy Ion Collisions}",
    eprint = "1703.00116",
    archivePrefix = "arXiv",
    primaryClass = "nucl-th",
    doi = "10.1103/PhysRevC.96.014914",
    journal = "Phys. Rev. C",
    volume = "96",
    number = "1",
    pages = "014914",
    year = "2017"
}

@article{Garcia-Montero:2023lrd,
    author = {Garcia-Montero, Oscar and Mazeliauskas, Aleksas and Plaschke, Philip and Schlichting, S{\"o}ren},
    title = "{Pre-equilibrium photons from the early stages of heavy-ion collisions}",
    eprint = "2308.09747",
    archivePrefix = "arXiv",
    primaryClass = "hep-ph",
    doi = "10.1007/JHEP03(2024)053",
    journal = "JHEP",
    volume = "03",
    pages = "053",
    year = "2024"
}

@article{Xu:2004mz,
    author = "Xu, Zhe and Greiner, Carsten",
    title = "{Thermalization of gluons in ultrarelativistic heavy ion collisions by including three-body interactions in a parton cascade}",
    eprint = "hep-ph/0406278",
    archivePrefix = "arXiv",
    doi = "10.1103/PhysRevC.71.064901",
    journal = "Phys. Rev. C",
    volume = "71",
    pages = "064901",
    year = "2005"
}

@article{Kasmaei:2019ofu,
    author = "Kasmaei, Babak Salehi and Strickland, Michael",
    title = "{Photon production and elliptic flow from a momentum-anisotropic quark-gluon plasma}",
    eprint = "1911.03370",
    archivePrefix = "arXiv",
    primaryClass = "hep-ph",
    doi = "10.1103/PhysRevD.102.014037",
    journal = "Phys. Rev. D",
    volume = "102",
    number = "1",
    pages = "014037",
    year = "2020"
}

@article{Gale:2014dfa,
    author = "Gale, Charles and Hidaka, Yoshimasa and Jeon, Sangyong and Lin, Shu and Paquet, Jean-Fran{\c{c}}ois and Pisarski, Robert D. and Satow, Daisuke and Skokov, Vladimir V. and Vujanovic, Gojko",
    title = "{Production and Elliptic Flow of Dileptons and Photons in a Matrix Model of the Quark-Gluon Plasma}",
    eprint = "1409.4778",
    archivePrefix = "arXiv",
    primaryClass = "hep-ph",
    reportNumber = "BNL-105041-2014-JA, RBRC-1093, RIKEN-QHP-151",
    doi = "10.1103/PhysRevLett.114.072301",
    journal = "Phys. Rev. Lett.",
    volume = "114",
    pages = "072301",
    year = "2015"
}

@article{Lee:2014pwa,
    author = "Lee, Chang-Hwan and Zahed, Ismail",
    title = "{Electromagnetic Radiation in Hot QCD Matter: Rates, Electric Conductivity, Flavor Susceptibility and Diffusion}",
    eprint = "1403.1632",
    archivePrefix = "arXiv",
    primaryClass = "hep-ph",
    doi = "10.1103/PhysRevC.90.025204",
    journal = "Phys. Rev. C",
    volume = "90",
    number = "2",
    pages = "025204",
    year = "2014"
}

@article{Tuchin:2014iua,
    author = "Tuchin, Kirill",
    title = "{Electromagnetic field and the chiral magnetic effect in the quark-gluon plasma}",
    eprint = "1411.1363",
    archivePrefix = "arXiv",
    primaryClass = "hep-ph",
    doi = "10.1103/PhysRevC.91.064902",
    journal = "Phys. Rev. C",
    volume = "91",
    number = "6",
    pages = "064902",
    year = "2015"
}

@article{Wang:2022jxx,
    author = "Wang, Xinyang and Shovkovy, Igor A.",
    title = "{Rate and ellipticity of dilepton production in a magnetized quark-gluon plasma}",
    eprint = "2205.00276",
    archivePrefix = "arXiv",
    primaryClass = "nucl-th",
    doi = "10.1103/PhysRevD.106.036014",
    journal = "Phys. Rev. D",
    volume = "106",
    number = "3",
    pages = "036014",
    year = "2022"
}

@article{Wang:2020dsr,
    author = "Wang, Xinyang and Shovkovy, Igor A. and Yu, Lang and Huang, Mei",
    title = "{Ellipticity of photon emission from strongly magnetized hot QCD plasma}",
    eprint = "2006.16254",
    archivePrefix = "arXiv",
    primaryClass = "hep-ph",
    doi = "10.1103/PhysRevD.102.076010",
    journal = "Phys. Rev. D",
    volume = "102",
    number = "7",
    pages = "076010",
    year = "2020"
}

@article{Basar:2012bp,
    author = "Basar, Gokce and Kharzeev, Dmitri and Kharzeev, Dmitri and Skokov, Vladimir",
    title = "{Conformal anomaly as a source of soft photons in heavy ion collisions}",
    eprint = "1206.1334",
    archivePrefix = "arXiv",
    primaryClass = "hep-ph",
    reportNumber = "BNL-98121-2012-JA",
    doi = "10.1103/PhysRevLett.109.202303",
    journal = "Phys. Rev. Lett.",
    volume = "109",
    pages = "202303",
    year = "2012"
}

@article{Sun:2024vsz,
    author = "Sun, Jing{\textquoteright}An and Yan, Li",
    title = "{The Weak Magnetic Photon Emission from Quark-gluon Plasma}",
    doi = "10.11804/NuclPhysRev.41.2023CNPC14",
    journal = "Nucl. Phys. Rev.",
    volume = "41",
    number = "1",
    pages = "558--563",
    year = "2024"
}

@article{Muller:2013ila,
    author = "Muller, Berndt and Wu, Shang-Yu and Yang, Di-Lun",
    title = "{Elliptic flow from thermal photons with magnetic field in holography}",
    eprint = "1308.6568",
    archivePrefix = "arXiv",
    primaryClass = "hep-th",
    doi = "10.1103/PhysRevD.89.026013",
    journal = "Phys. Rev. D",
    volume = "89",
    number = "2",
    pages = "026013",
    year = "2014"
}

@article{Arciniega:2013dqa,
    author = "Arciniega, Gustavo and Nettel, Francisco and Ortega, Patricia and Pati{\~n}o, Leonardo",
    title = "{Brighter Branes, enhancement of photon production by strong magnetic fields in the gauge/gravity correspondence}",
    eprint = "1307.1153",
    archivePrefix = "arXiv",
    primaryClass = "hep-th",
    doi = "10.1007/JHEP04(2014)192",
    journal = "JHEP",
    volume = "04",
    pages = "192",
    year = "2014"
}

@article{Avila:2022cpa,
    author = "{\'A}vila, Daniel and Nettel, Francisco and Pati{\~n}o, Leonardo",
    title = "{Darker and brighter branes: Suppression and enhancement of photon production in a strongly coupled magnetized plasma}",
    eprint = "2204.00024",
    archivePrefix = "arXiv",
    primaryClass = "hep-th",
    doi = "10.1103/PhysRevD.107.066010",
    journal = "Phys. Rev. D",
    volume = "107",
    number = "6",
    pages = "066010",
    year = "2023"
}

@article{Castano-Yepes:2024vlj,
    author = "Casta{\~n}o-Yepes, Jorge David and Mu{\~n}oz, Enrique",
    title = "{Anisotropic photon and dilepton yield in a thermalized quark-gluon plasma under spatial magnetic fluctuations}",
    eprint = "2412.14055",
    archivePrefix = "arXiv",
    primaryClass = "hep-th",
    doi = "10.1103/PhysRevD.111.076028",
    journal = "Phys. Rev. D",
    volume = "111",
    number = "7",
    pages = "076028",
    year = "2025"
}

@article{Ayala:2017vex,
    author = "Ayala, Alejandro and Castano-Yepes, Jorge David and Dominguez, Cesareo A. and Hernandez, Luis A. and Hernandez-Ortiz, Saul and Tejeda-Yeomans, Maria Elena",
    title = "{Prompt photon yield and elliptic flow from gluon fusion induced by magnetic fields in relativistic heavy-ion collisions}",
    eprint = "1704.02433",
    archivePrefix = "arXiv",
    primaryClass = "hep-ph",
    doi = "10.1103/PhysRevD.96.014023",
    journal = "Phys. Rev. D",
    volume = "96",
    number = "1",
    pages = "014023",
    year = "2017",
    note = "[Erratum: Phys.Rev.D 96, 119901 (2017)]"
}

@article{Ayala:2019jey,
    author = "Ayala, Alejandro and Casta{\~n}o-Yepes, Jorge David and Dominguez Jimenez, Isabel and Salinas San Mart{\'\i}n, Jordi and Tejeda-Yeomans, Mar{\'\i}a Elena",
    title = "{Centrality dependence of photon yield and elliptic flow from gluon fusion and splitting induced by magnetic fields in relativistic heavy-ion collisions}",
    eprint = "1904.02938",
    archivePrefix = "arXiv",
    primaryClass = "hep-ph",
    doi = "10.1140/epja/s10050-020-00060-9",
    journal = "Eur. Phys. J. A",
    volume = "56",
    number = "2",
    pages = "53",
    year = "2020"
}

@article{Ayala:2022zhu,
    author = "Ayala, Alejandro and Casta{\~n}o-Yepes, Jorge David and Hern{\'a}ndez, L. A. and Mizher, Ana Julia and Tejeda-Yeomans, Mar{\'\i}a Elena and Zamora, R.",
    title = "{Anisotropic photon emission from gluon fusion and splitting in a strong magnetic background: The two-gluon one-photon vertex}",
    eprint = "2209.09364",
    archivePrefix = "arXiv",
    primaryClass = "hep-ph",
    doi = "10.1103/PhysRevC.106.064905",
    journal = "Phys. Rev. C",
    volume = "106",
    number = "6",
    pages = "064905",
    year = "2022"
}

@article{Ayala:2024ucr,
    author = "Ayala, Alejandro and Bernal-Langarica, Santiago and Jaber-Urquiza, Jorge and Medina-Serna, Jos{\'e} Jorge",
    title = "{Two-gluon one-photon vertex in a magnetic field and its explicit one-loop approximation in the intermediate field strength regime}",
    eprint = "2406.18673",
    archivePrefix = "arXiv",
    primaryClass = "hep-ph",
    doi = "10.1103/PhysRevD.110.076021",
    journal = "Phys. Rev. D",
    volume = "110",
    number = "7",
    pages = "076021",
    year = "2024"
}

@article{Baier:2000sb,
    author = "Baier, R. and Mueller, Alfred H. and Schiff, D. and Son, D. T.",
    title = "{'Bottom up' thermalization in heavy ion collisions}",
    eprint = "hep-ph/0009237",
    archivePrefix = "arXiv",
    doi = "10.1016/S0370-2693(01)00191-5",
    journal = "Phys. Lett. B",
    volume = "502",
    pages = "51--58",
    year = "2001"
}

@article{Garcia-Montero:2019vju,
    author = "Garcia-Montero, Oscar",
    title = "{Non-equilibrium photons from the bottom-up thermalization scenario}",
    eprint = "1909.12294",
    archivePrefix = "arXiv",
    primaryClass = "hep-ph",
    doi = "10.1016/j.aop.2022.168984",
    journal = "Annals Phys.",
    volume = "443",
    pages = "168984",
    year = "2022"
}

@article{Monnai:2019vup,
    author = "Monnai, Akihiko",
    title = "{Prompt, pre-equilibrium, and thermal photons in relativistic nuclear collisions}",
    eprint = "1907.09266",
    archivePrefix = "arXiv",
    primaryClass = "nucl-th",
    reportNumber = "KEK-TH-2144",
    doi = "10.1088/1361-6471/ab8d8c",
    journal = "J. Phys. G",
    volume = "47",
    number = "7",
    pages = "075105",
    year = "2020"
}

@article{Papanyan:1972cv,
    author = "Papanyan, V. O. and Ritus, V. I.",
    title = "{Vacuum polarization and photon splitting in an intense field}",
    journal = "Zh. Eksp. Teor. Fiz.",
    volume = "61",
    pages = "2231--2241",
    year = "1972"
}

@article{Papanyan:1974xa,
    author = "Papanyan, V. O. and Ritus, V. I.",
    title = "{The three-photon interaction in intense fields and scale invariance}",
    journal = "Zh. Eksp. Teor. Fiz.",
    volume = "65",
    number = "5",
    pages = "1756--1771",
    year = "1974"
}

@article{PHENIX:2022rsx,
    author = "Abdulameer, N. J. and others",
    collaboration = "PHENIX",
    title = "{Nonprompt direct-photon production in Au+Au collisions at sNN=200 GeV}",
    eprint = "2203.17187",
    archivePrefix = "arXiv",
    primaryClass = "nucl-ex",
    doi = "10.1103/PhysRevC.109.044912",
    journal = "Phys. Rev. C",
    volume = "109",
    number = "4",
    pages = "044912",
    year = "2024"
}

@article{Kurkela:2015qoa,
    author = "Kurkela, Aleksi and Zhu, Yan",
    title = "{Isotropization and hydrodynamization in weakly coupled heavy-ion collisions}",
    eprint = "1506.06647",
    archivePrefix = "arXiv",
    primaryClass = "hep-ph",
    reportNumber = "CERN-PH-TH-2015-142",
    doi = "10.1103/PhysRevLett.115.182301",
    journal = "Phys. Rev. Lett.",
    volume = "115",
    number = "18",
    pages = "182301",
    year = "2015"
}

@article{Ayala:2024jvc,
    author = "Ayala, Alejandro and Mizher, Ana Julia",
    title = "{Influence of magnetic field-induced anisotropic gluon pressure during pre-equilibrium in heavy-ion collisions: A faster road toward isotropization}",
    eprint = "2407.09754",
    archivePrefix = "arXiv",
    primaryClass = "hep-ph",
    doi = "10.1103/PhysRevD.110.L111501",
    journal = "Phys. Rev. D",
    volume = "110",
    number = "11",
    pages = "L111501",
    year = "2024"
}




\end{document}